\documentclass[runningheads]{llncs}

\usepackage[english]{babel}
\usepackage{graphicx}
\usepackage{mathtools}
\usepackage{amsfonts}
\usepackage{amsmath}
\usepackage{amssymb}
\usepackage{bm}
\usepackage{nicefrac}
\usepackage[separate-uncertainty = true]{siunitx}
\usepackage{booktabs}
\usepackage{makecell}
\usepackage{multirow}
\usepackage{mathtools}
\usepackage{dblfloatfix}
\usepackage{csquotes}
\usepackage{hyperref}
\usepackage[capitalise]{cleveref}
\usepackage{etoolbox}
\robustify\bfseries
\robustify\itshape
\begin{document}

\title{Continuous-Time Deep Glioma Growth Models}

\author{
Jens Petersen\inst{1} \and
Fabian Isensee\inst{2} \and
Gregor Köhler\inst{1}
Paul F. Jäger\inst{3} \and
David Zimmerer\inst{1} \and
Ulf Neuberger\inst{4} \and
Wolfgang Wick\inst{5,6} \and
Jürgen Debus\inst{7,8,9} \and
Sabine Heiland\inst{4} \and
Martin Bendszus\inst{4} \and
Philipp Vollmuth\inst{4} \and
Klaus H. Maier-Hein\inst{1}
}


\authorrunning{J. Petersen et al.}

\institute{
Division of Medical Image Computing, German Cancer Research Center, Heidelberg, Germany \and
HIP Applied Computer Vision Lab, Division of Medical Image Computing, German Cancer Research Center \and
Interactive Machine Learning Group, German Cancer Research Center \and
Department of Neuroradiology, Heidelberg University Hospital, Heidelberg, Germany \and
Neurology Clinic, Heidelberg University Hospital \and
DKTK CCU Neurooncology, German Cancer Research Center \and
Division of Molecular and Translational Radiation Oncology, Heidelberg Institute of Radiation Oncology (HIRO), Heidelberg, Germany \and
Heidelberg Ion-Beam Therapy Center (HIT), Heidelberg University Hospital \and
Clinical Cooperation Unit Radiation Oncology, German Cancer Research Center
\email{jens.petersen@dkfz.de}
}

\maketitle

\begin{abstract}

The ability to estimate how a tumor might evolve in the future could have tremendous clinical benefits, from improved treatment decisions to better dose distribution in radiation therapy.
Recent work has approached the glioma growth modeling problem via deep learning and variational inference, thus learning growth
dynamics entirely from a real patient data distribution. So far, this approach was constrained to predefined image acquisition intervals and sequences of fixed length, which limits its applicability in more realistic scenarios.
We overcome these limitations by extending Neural Processes, a class of conditional generative models for stochastic time series, with a hierarchical multi-scale representation encoding including a spatio-temporal attention mechanism. The result is a learned growth model that can be conditioned on an arbitrary number of observations, and that can produce a distribution of temporally consistent growth trajectories on a continuous time axis. On a dataset of 379 patients, the approach successfully captures both global and finer-grained variations in the images, exhibiting superior performance compared to other learned growth models. 

\end{abstract}

\section{Introduction}
\label{sec:introduction}

Glioma growth modeling refers to the problem of estimating how the tumor burden in a patient might evolve in the future given the current state of the disease and possibly its history. In the past, this was often approached with so-called reaction-diffusion models, where tumor growth is described by the physical diffusion of an assumed tumor cell density. There is a large body of related literature, so we only refer to \cite{mang_integrated_2020,menze_image-based_2011} as overviews. Newer works not included therein are \cite{subramanian_multiatlas_2020,ezhov_neural_2019}, among others.
All of these approaches have in common that they depend on the quality of the chosen forward model, which always includes simplifying assumptions when compared to the actual biological processes.

A recent alternative was proposed in \cite{petersen_deep_2019} in the form of fully learned growth models. The authors used a probabilistic U-Net to map a fixed-length input sequence of MRI scans to a distribution of possible spatial tumor configurations in the future. The model thus learns a growth model directly from data and doesn't rely on explicit assumptions and simplifications with respect to the underlying biological process. At the same time, the approach suffers from several limitations, requiring a fixed number of input observations, requiring fixed time intervals between observations, and being able to make predictions only for a single predefined future time point. In order to overcome all of these limitations, we build upon the Neural Process framework \cite{garnelo_conditional_2018,garnelo_neural_2018}, a family of conditional generative models specifically conceived to model continuous-valued stochastic processes. They work by encoding so-called \emph{context} observations into a joint representation and then predicting a desired \emph{target} output from it. In particular, we extend the work from \cite{eslami_neural_2018,kumar_consistent_2018}, who apply Neural Processes in the image domain, by introducing a convolutional encoder-decoder structure with representations at multiple scales. The single-scale encoder-decoder setups used in \cite{eslami_neural_2018,kumar_consistent_2018} require extremely large decoders to produce outputs with sufficient spatial resolution. As a result, they require vast amounts of data to be trained successfully, and the authors only demonstrate their approaches on synthetic images. Ours is one of the first Neural Processes to be applied to real data in the image domain\footnote{\emph{Image domain} refers to the fact that the observations are entire images. The original Neural Processes \cite{garnelo_conditional_2018,garnelo_neural_2018} work on images by treating individual pixels as observations.}, with \cite{kia_neural_2019} the only other work we're aware of. To allow our approach to model tumor changes with high spatial resolution, we also introduce an attention mechanism \cite{vaswani_attention_2017} that aggregates the context information at different scales. In contrast to existing work with attention in Neural Processes \cite{kim_attentive_2019,rosenbaum_learning_2018}, we design the mechanism so that it can model spatial and temporal dependencies simultaneously.

In summary, this work makes the following contributions: 1.) We propose the first fully learned tumor growth model that can make predictions at arbitrary points on the time axis, conditioned on an arbitrary number of available observations. 2.) In doing so, we propose a Neural Process variant for image-based problems that is efficient enough to work successfully on real-world data. 3.) We show that the proposed approach significantly outperforms prior work on the task of learning-based modeling of glioma growth. We compare our approach with both the model proposed in \cite{petersen_deep_2019} and a naive convolutional Neural Process. 4.) We provide a complete implementation at \url{https://github.com/MIC-DKFZ/deep-glioma-growth}.

\begin{figure}[t]
    \centering
    \includegraphics[width=0.87\textwidth]{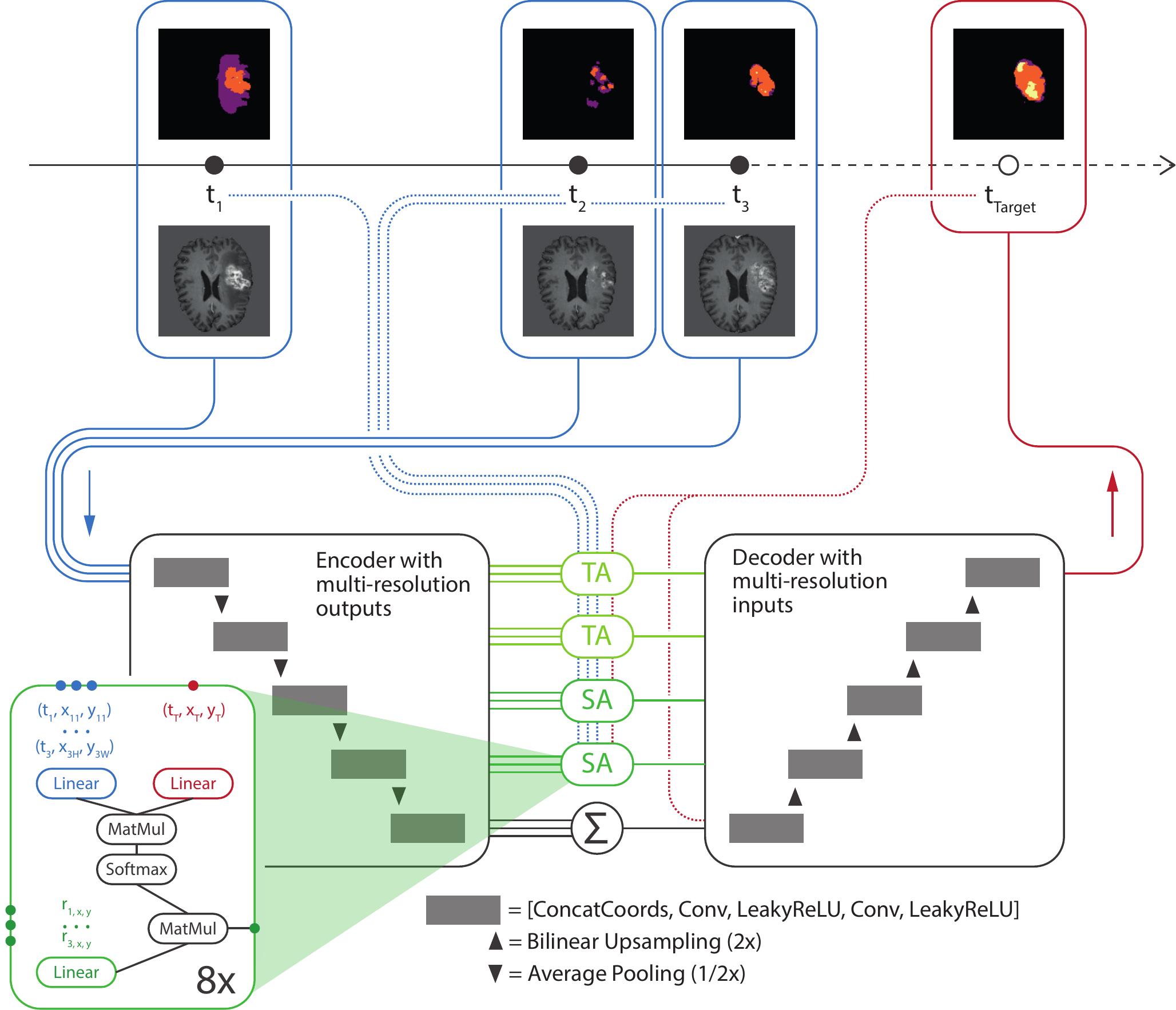}
    \caption{A schematic of the proposed model. All \emph{context} observations (MRI/segmentation/time, blue) are encoded separately with the same convolutional encoder that outputs representations at different resolution scales. The lowest scale (black) has no spatial resolution, and the corresponding representations are summed. The result is a global distribution, a sample of which is fed to the decoder together with the \emph{target} input (time, red). For representations with low spatial resolution, we use spatio-temporal attention (SA) to aggregate context information. At higher resolutions we use only temporal attention (TA) to reduce GPU memory requirements. ConcatCoords represents an operator that appends pixel locations (\num{-0.5} to \num{0.5} per axis). The first encoder block also uses InstanceNorm.}
    \label{fig:method}
\end{figure}

\section{Methods}
\label{sec:methods}

\subsection{Model}
\label{sub:model}

The model we propose is shown in \cref{fig:method}. We have a number of observations (MRIs and corresponding tumor segmentations) available at continuous time values $\bm{t_c}$. We call those the \emph{context}, with bold indicating a collection henceforth. We are interested in possible tumor segmentations for future \emph{target} times $\bm{t_{t}}$. Each context observation is encoded separately with an encoder that outputs representations hierarchically at multiple scales: $\bm{r_c=(r_{c,32},r_{c,16},r_{c,8},r_{c,4},r_{c,global})}$. The second index refers to the scale (the first entry has a resolution of $32^2$, etc.) and the global representations $\bm{r_{c,global}}$ describe mean and standard deviation of diagonal normal distributions. These means and standard deviations are summed over all context encodings (bottom path in \cref{fig:method}) to form a single global distribution. Because it is independent of the target $t_t$, this allows us to sample growth trajectories that are consistent over time and that can be evaluated at arbitrary new points in time. A sample from the distribution (which is 128-dimensional in our case) along with $t_{t}$ is fed to the decoder to predict the corresponding segmentation. If only the global representation is used, our model reduces to a Neural Process \cite{garnelo_conditional_2018,garnelo_neural_2018} with a convolutional encoder and decoder. We want our approach to be able to model spatial and temporal dependencies simultaneously, so we aggregate the context information at higher scales using a spatio-temporal dot-product attention mechanism \cite{vaswani_attention_2017}, conditioned on the target $t_{target}$ and pixel locations $(x, y)$ at the given scale (see inset in \cref{fig:method}):

\begin{gather}
    r(\bm{t_t},\bm{x_t},\bm{y_t}) = \mathrm{softmax}\left(\frac{\bm{QK}^T}{\sqrt{d}}\right)\bm{V}\\
    \bm{Q}=\mathrm{Lin}\,(\bm{t_t},\bm{x_t},\bm{y_t},\bm{r_{c,x,y}})\;,\;
    \bm{K}=\mathrm{Lin}\,(\bm{t_c},\bm{x_c},\bm{y_c},\bm{r_{c,x,y}})\;,\;
    \bm{V}=\mathrm{Lin}\,(\bm{r_{c,x,y}})
\end{gather}

$\bm{Q},\bm{K},\bm{V}$ are linear maps from what is usually referred to as \emph{queries} (target locations $\bm{(t_t,x_t,y_t)}$), \emph{keys} (available locations $\bm{(t_c,x_c,y_c)}$) and \emph{values} (the representations at the \emph{key} locations). We also use $\bm{r_{c,x,y}}$ to construct $\bm{Q}$ and $\bm{K}$ because of the success of self-attention in transformer architectures, e.g. \cite{devlin_bert:_2018}. This means that the available representations are weighted according to the similarity of the corresponding key to the queries. The similarity is given by the dot product in a higher-dimensional space (with dimension $d=16$ in our case). Note that in \cref{fig:method}, each attention block represents so-called multihead attention, meaning 8 parallel blocks of the above. The downside of this spatio-temporal attention is that the matrix product $\bm{QK^T}$ requires GPU memory that scales quadratically with the number of pixels. As a result we only use the full spatio-temporal attention at the two lowest scales (resolutions $4^2$ and $8^2$) and resort to temporal attention at higher scales, i.e. the same as above but without a dependence on pixel locations. 

Our model is trained with the following optimization objective, where $I$ are inputs (MRI scans) while $S$ refers to segmentations. During training the context is absorbed into the target set ($\bm{t_t}\rightarrow\bm{t_{t+c}}$, etc.), meaning we let the models reconstruct the inputs as well instead of only predicting a future segmentation:

\begin{align}
    \min_\theta\mathop{\mathbb{E}}_{z\sim q_\theta(z|\bm{I_t},\bm{S_t}, \bm{t_t})}&-\log p_\theta(\bm{S_t}|\bm{I_c},\bm{S_c},\bm{t_c},\bm{t_t},z)\nonumber\\
    &+\beta\cdot D_{KL}(q_\theta(z|\bm{I_t},\bm{S_t},\bm{t_t})||q_\theta(z|\bm{I_c},\bm{S_c},\bm{t_c}))
    \label{eq:optimization_variational}
\end{align}

where $q_\theta(z|\cdot)$ is the global representation predicted by the encoder for the given inputs. This is the Neural Process variant of the well-known variational inference objective, with some modifications: 1.) There is a direct dependence of the reconstruction likelihood $p_\theta$ on $(\bm{I_c,S_c,t_c})$, because the attention mechanisms are deterministic (otherwise the influence would only be via $z$). 2.) The predictive likelihood is usually modeled as a factorizing categorical likelihood, leading to a cross-entropy loss. We instead use the sum of cross-entropy and Dice losses \cite{drozdzal_importance_2016}. As a result, the likelihood no longer factorizes over pixels, so 3.) we introduce a weight $\beta$ for the KL term, as done for example in \cite{higgins_beta-vae:_2016}, which re-balances the two loss terms. We optimized this parameter through line-searching the space of possible values in powers of 10, choosing the one that lead to KL loss values most similar to those obtained during training with only cross-entropy: $\beta=0.0001$. Note that the objective for the baseline from \cite{petersen_deep_2019} can be written the same way, but without dependence on $(\bm{S_c,t_c,t_t})$. Models are trained for 300 epochs with Adam \cite{kingma_adam_2015} and a learning rate of \num{0.0001} at batch size 128. We give between 2 and 5 inputs as context and always have a single target point, but the models predict both context and target during training. For all other hyperparameters we refer to the provided implementation. 

\subsection{Data \& Evaluation}
\label{sub:dataeval}

We use a dataset that consists of MRI scans from 379 glioblastoma patients, stemming from a multi-center study that compared treatment using a combination of chemotherapy (Lomustine) and an angiogenesis inhibitor (Bevacizumab) with treatment using chemotherapy alone \cite{wick_lomustine_2017}. The study found no significant difference between test and control arm in terms of overall survival, so we ignore treatment effects in our work and treat it as an additional source of stochasticity. Pre-processing of the data including annotations (segmentations of edema, enhancing tumor and necrosis) was performed by experienced neuroradiologists as described previously \cite{kickingereder_automated_2019}. For each patient there are between 3 and 13 (mean \num{4.85}) longitudinal scans, consisting of native T1, contrast-enhanced T1 (with gadolinium agent), T2 and FLAIR. All scans for a patient are skull-stripped, registered to the native T1 image of the first available scan, and normalized by mean and standard deviation on a per-channel basis. We resample the original data to isotropic $128^3$ and then extract a region of size $64^3$ centered around the tumor. We find that this offers a good trade-off between resolution and compute requirements. Finally, we work on axial slices of the resulting arrays, making extensive use of data augmentation during training. A full list of transforms is given in the implementation. Note that our model only uses skip connections with a resolution of at most $32^2$. For the evaluation we perform a random 80/20 train/test split of the data (stratified at the patient level to avoid data leakage). Hyperparameter optimization is done via cross-validation on the training set, for the final results we train all models on the full training set and report scores on the test set, leaving out those instances where the true Dice overlap between the target point and the last context point is 0.

We compare our approach with the one presented in \cite{petersen_deep_2019} using their official implementation and a hyperparameter configuration that matches our model (U-Net layer size and depth, loss, etc.). We further compare with a regular Neural Process \cite{garnelo_neural_2018}, which is a special case of our model where all the attention skip-connections are removed (see \cref{fig:method}). Because the approach in \cite{petersen_deep_2019} works with a fixed number of input timesteps, we train 4 separate models (2-5 input timesteps) and evaluate them jointly. We look at three different metrics: 1) The \emph{Test Loss}, which is just the optimization objective (\cref{eq:optimization_variational}) evaluated on the test set, but only using the test points and not the context reconstructions. It is by definition the optimal measure of how well the models perform the task they are trained to perform. Note that this measure is only meaningful when all models are trained with the same loss, which we do. 2) The \emph{Surprise} is the KL divergence, i.e. the second summand in \cref{eq:optimization_variational}. It can be interpreted as a measure of how much the model has to adjust its prior belief when presented with the true future observation (the posterior). 3) The \emph{Query Volume Dice} takes 100 random samples from the predicted prior and measures the Dice overlap for the sample that best matches the observed future in terms of whole tumor volume. It is best interpreted from an application perspective, when clinicians are interested in possible spatial growth estimates \emph{conditioned} on a potential volume increase. Note that we are mainly interested in a comparison with other learned growth models, and the chosen metrics are not suitable for a comparison with diffusion-based models. We leave such an evaluation for future work.

\begin{figure}[t]
    \centering
    \includegraphics[width=0.78\textwidth]{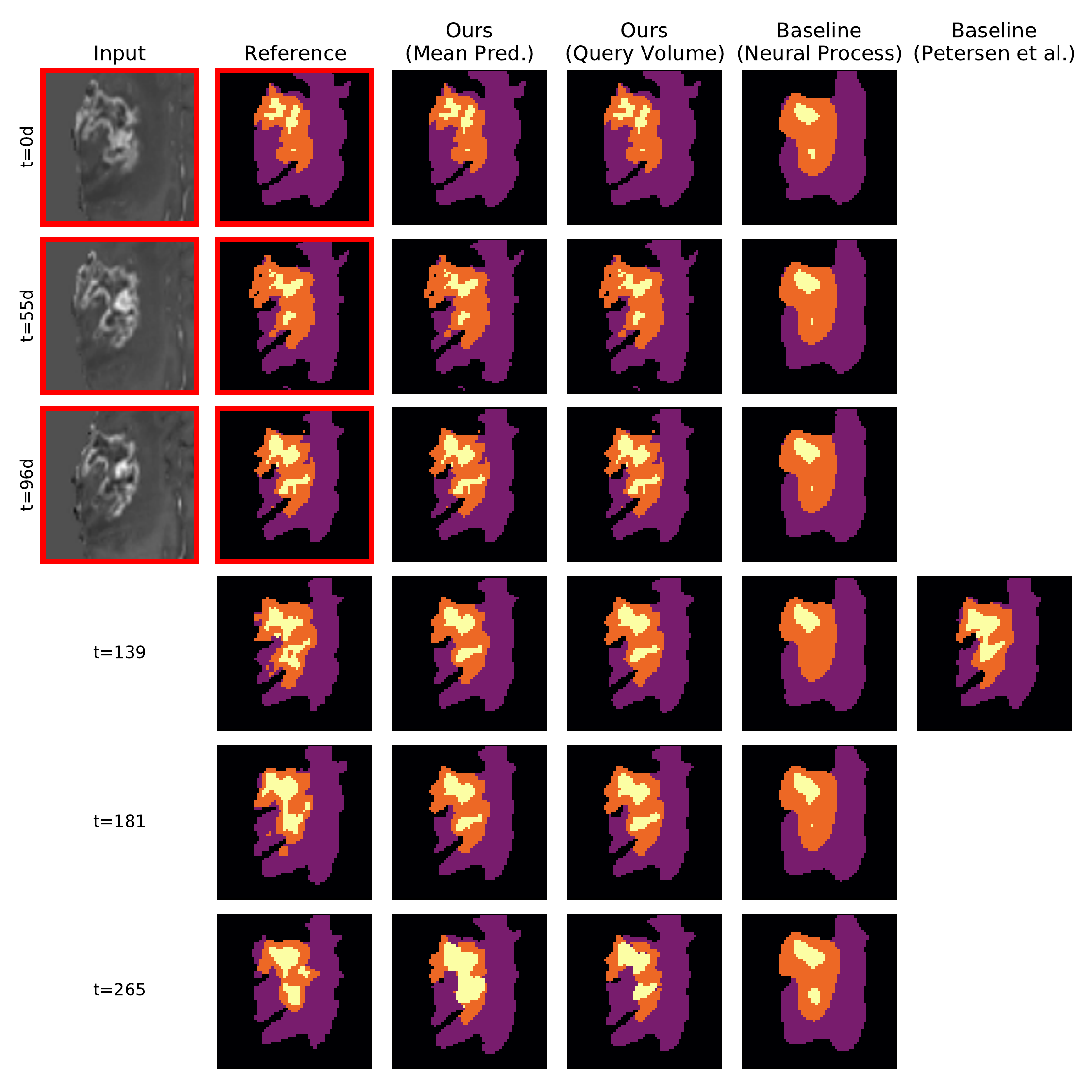}
    \caption{Example case with predictions from all methods we compare in this work (purple=edema, orange=enhancing tumor, yellow=necrosis). Red borders mark the context, we only show the T1c input channel. Our model can reconstruct the context observations very accurately, while the Neural Process prediction lacks spatial resolution. The mean prediction of our model anticipates an expansion of the necrosis in the tumor center (at $t=265$ days), which is commonly observed, but does not occur in this particular instance. The query volume prediction is very close to the observed future, indicating that our model does indeed learn meaningful growth trajectories. The model from \cite{petersen_deep_2019} can only make a prediction at a single fixed point in time.}
    \label{fig:example}
\end{figure}

\begin{table}[t]

    \centering
    
    \caption{Results on the test set. $\uparrow/\downarrow$ indicate that higher/lower is better, bold marks the best-performing method. Errors represent the standard error of the mean. \emph{Test Loss} evaluates \cref{eq:optimization_variational} on the test set, \emph{Surprise} measures the distance between prior and posterior distributions. \emph{Query Volume Dice} draws 100 samples from the prior and evaluates the Dice overlap for the sample that is closest to the true future segmentation in terms of tumor volume. Our model outperforms the baselines in all metrics.}
    \label{tab:results}
    
    \sisetup{detect-all}
    \begin{tabular}{lS[table-format=2.1]S[table-format=2.1(2)]S[table-format=3.1(2)]S[table-format=2.1(2)]}
    \toprule
    &  & \multicolumn{1}{c}{Test Loss $\downarrow$} & \multicolumn{1}{c}{Surprise $\downarrow$} & \multicolumn{1}{c}{Query Vol. Dice $\uparrow$} \\
    & \multicolumn{1}{c}{Parameters} & \multicolumn{1}{c}{[1e-2]} & \multicolumn{1}{c}{[nats]} & \multicolumn{1}{c}{[1e-2]} \\
    \midrule
    Neural Process \cite{garnelo_neural_2018} & 5.9M & 34.1(2) & 431.9(37) & 57.5(2) \\
    Learned Discrete \cite{petersen_deep_2019} & 11.2M & 38.9(3) & 112.1(7) & 60.1(2) \\
    Learned Continuous (ours) & 6.4M & \bfseries 24.1(2) & \bfseries 82.7(15) & \bfseries 71.4(2) \\
    \bottomrule
    \end{tabular}
    
\end{table}

\begin{figure}[t]
    \centering
    \includegraphics[width=\textwidth]{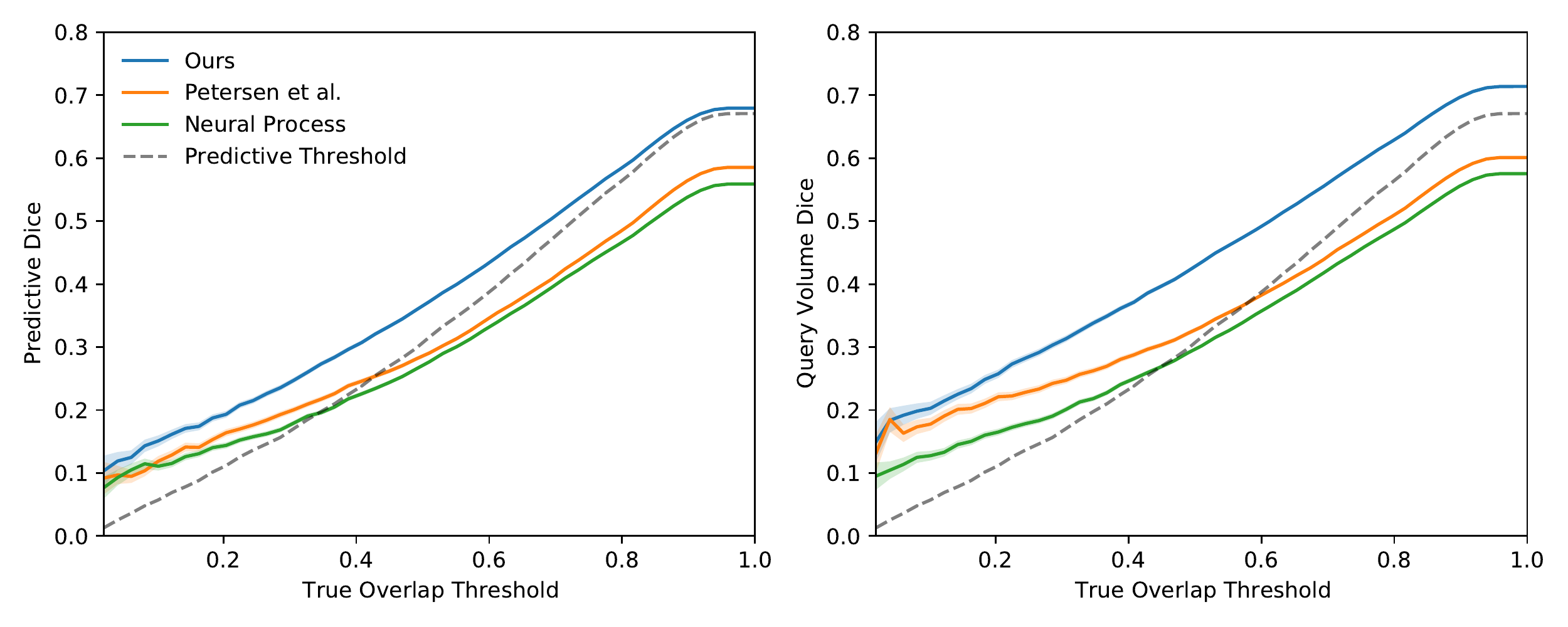}
    \caption{\emph{Predictive Dice} and \emph{Query Volume Dice} at different thresholds of true overlap. There is often very little change between consecutive observations, so we filter test cases by varying a threshold of their true overlap (Dice) to see how performance changes. As an example, if we select a threshold of 0.6, the corresponding y-values are the average performances on test cases where the overlap of the test timestep with the last context input is less than 0.6. The \emph{predictive threshold} is the average true overlap of those thresholded cases (and the average true Dice for all test cases is \num{0.67}), meaning that if a model falls below that line, it would be better to just assume no change will take place instead of trusting the model's prediction. Only our proposed model stays above that line for all thresholds.}
    \label{fig:dice_vs_threshold}
\end{figure}

\section{Results}
\label{sec:results}

An exemplary case is shown in \cref{fig:example}, with predictions from all models. We find that our model produces predictions with high spatial resolution. It can reconstruct the context segmentations very accurately, while the Neural Process is unable to create sufficient spatial detail. The mean prediction from our model overestimates how fast the necrosis (yellow in \cref{fig:example}) in the tumor center expands, but it is important to consider that the observed future isn't necessarily the most likely one, so it is expected to find such differences. More importantly, when we sample from the predicted prior and take the samples that are closest to the observed future in terms of tumor volume, our model matches the observed future closely. The model from \cite{petersen_deep_2019} can produce spatial detail comparable to our model, but it can only make a prediction at a single fixed time relative to the context observations.

The average performance of all methods on the test set is shown in \cref{tab:results}. We evaluate the \emph{Test Loss} (i.e. \cref{eq:optimization_variational}, but only on target points, not context reconstructions), the \emph{Surprise}, meaning the KL divergence between prior and posterior, as well as the \emph{Query Volume Dice}, which is the Dice for the sample that best matches the observed future in terms of tumor volume (out of 100 samples from the prior). Our method significantly outperforms the competing approaches in all metrics, with only a little more than half the number of parameters that the model from \cite{petersen_deep_2019} uses, because ours doesn't require separate prior and posterior networks. Interestingly, the test loss of the Neural Process \cite{garnelo_neural_2018} is actually lower than that of the learned discrete model \cite{petersen_deep_2019}, even though its perceived quality is much lower (see \cref{fig:example}). It should be noted that the numerical results are not comparable to those reported in \cite{petersen_deep_2019}, because we use a slightly different training objective. There are many test cases with only very small changes between the test timestep and the previous context observation. To understand how the performance depends on changes that take place between those times, we iterate a threshold over their true Dice overlap to evaluate the average \emph{Predictive Dice} and Query Volume Dice on the subset of cases below the given threshold. The result is displayed in \cref{fig:dice_vs_threshold}, where we also include a predictive threshold, which is the average true overlap below the given threshold. Below that line, a model performs worse than one that simply predicts no change between consecutive times. Ours is the only approach that is always above the predictive threshold, while the others actually fall \emph{below} it at higher cutoffs.

\section{Discussion}
\label{sec:discussion}

This work proposes a fully learned glioma growth model, first introduced in \cite{petersen_deep_2019} as an alternative to commonly used biological diffusion models \cite{mang_integrated_2020,menze_image-based_2011}. While successful, the approach from \cite{petersen_deep_2019} has a number of practical limitations: it requires a fixed number of context observations, it requires a fixed time interval between consecutive observations and it can only make a prediction one time interval step into the future. Our proposed model overcomes all of those limitations and can be conditioned on arbitrarily many context observations on a continuous time axis. From these context observations, our model predicts a distribution over growth trajectories, where each sample is temporally consistent and can be evaluated at any desired continuous-valued time. Our model also significantly outperforms the one from \cite{petersen_deep_2019} in several metrics, which we demonstrate on a dataset ten times larger than the one used in \cite{petersen_deep_2019}. Our model's main limitation is the high GPU memory requirement of the spatio-temporal attention mechanism we introduce. This is a problem many attention and transformer architectures suffer from, and ways to make them less resource intensive are actively being researched \cite{kitaev_reformer_2020,wang_linformer_2020}. It's also the reason why we performed our experiments on two-dimensional slices rather than full 3D volumes. As a result, one should be careful not to draw any premature conclusions with respect to a possible clinical application. While our results look promising, the value of our model for clinical purposes, for example in radiation therapy, must be validated extensively. We leave such efforts for future work. A comparison with diffusion-based models, particularly in a radiation therapy context \cite{le_personalized_2017,lipkova_personalized_2019}, is another interesting opportunity for future work. Building on the Neural Process framework \cite{garnelo_conditional_2018,garnelo_neural_2018}, our proposed approach constitutes an efficient Neural Process variant for image time series and is, to the best of our knowledge, only the second time Neural Processes have been demonstrated on real data in the image domain \cite{kia_neural_2019}. We believe it can prove useful for both other types of tumor growth as well as any other kind of stochastic time series with image data.

\section*{Acknowledgements}
Part of this work was funded by the Helmholtz Imaging Platform (HIP), a platform of the Helmholtz Incubator on Information and Data Science.

\bibliographystyle{splncs04}
\bibliography{paper366}

\end{document}